\documentclass[10pt,journal,twocolumn]{IEEEtran}
\usepackage{cite}
\usepackage{amsmath,amssymb,amsfonts}
\usepackage{algorithmic}
\usepackage{graphicx}
\usepackage{textcomp}
\usepackage{blindtext}
\usepackage{setspace}
\usepackage{amstext}
\usepackage{amsmath}
\usepackage{amsfonts,amssymb}
\usepackage{amssymb}
\usepackage[dvips]{epsfig}
\usepackage{epsfig}
\usepackage[dvips]{graphicx}
\usepackage{multirow}
\usepackage[utf8]{inputenc}
\usepackage[english]{babel}
\usepackage{caption}
\usepackage{subcaption}
\usepackage{float}
\usepackage{adjustbox}
\usepackage{gensymb}
\usepackage[dvips]{epsfig}
\usepackage{epsfig}
\usepackage[dvips]{graphicx}
\usepackage{multirow}
\usepackage{multirow} 
\usepackage[utf8]{inputenc}
\usepackage[english]{babel}
\usepackage{mathtools,xparse}
\newtheorem{theorem}{Theorem}[section]

\newtheorem{lemma}[theorem]{Lemma}
\providecommand{\keywords}[1]{\textbf{\textit{Index terms---}} #1}
\def\by{{\boldsymbol{y}}}
\def\bx{{\boldsymbol{x}}}
\def\br{{\boldsymbol{r}}}
\def\bw{{\boldsymbol{w}}}
\def\bQ{{\boldsymbol{Q}}}
\def\bM{{\boldsymbol{M}}}

\def\bA{{\boldsymbol{A}}}
\def\bb{{\boldsymbol{b}}}

\def\bJ{{\boldsymbol{J}}}
\def\bH{{\boldsymbol{H}}}
\def\bh{{\boldsymbol{h}}}

\def\bp{{\boldsymbol{p}}}

\def\bu{{\boldsymbol{u}}}
\def\bv{{\boldsymbol{v}}}
\def\bd{{\boldsymbol{d}}}
\def\bz{{\boldsymbol{z}}}
\def\bg{{\boldsymbol{g}}}

\begin{document}
\title{SOLVIT: A Reference–Free Source Localization Technique using Majorization Minimization}
\author{R.~Jyothi and P.~Babu
\thanks{The authors are with CARE, IIT Delhi, New Delhi, 110016, India.(email: jyothi.r@care.iitd.ac.in, prabhubabu@care.iitd.ac.in}}
\maketitle 
\begin{abstract}
We consider the problem of localizing the source using range and range-difference measurements. Both the problems are non-convex and non-smooth and are challenging to solve. In this paper, we develop an iterative algorithm - \textbf{So}urce \textbf{L}ocalization \textbf{V}ia an \textbf{It}erative technique (\textbf{SOLVIT}) to localize the source using all the distinct range-difference measurements, i.e., without choosing a reference sensor. \textbf{SOLVIT} is based on the Majorization Minimization approach - in which a novel upper bound is formulated and minimized to get a closed-form solution at every iteration. We also solve the source localization problem based on range measurements and rederive the Standard Fixed Point algorithm using the Majorization Minimization approach. By doing so, we show a less intricate way to prove the convergence of the Standard Fixed Point algorithm. Numerical simulations and experiments in an anechoic chamber confirm that \textbf{SOLVIT} performs better than existing reference-based and reference-free methods in terms of source positioning accuracy. 
\end{abstract}
\section{Introduction} \label{sec:1}
The problem of localizing a source using an array of sensors has received significant attention because of the problems applications in teleconferencing, radar, sonar and wireless communications \cite{mle_suboptimal}, \cite{stoica}, and \cite{teleconference}. Depending on the nature of the source, its position can be estimated by either using active or passive source localization technique. In the case of active source localization, each sensor transmits a signal towards an object of interest and receives the reflected signal from which the range $r_{i}$ i.e., the distance between the source and the $i$th sensor is estimated. To localize a passive source, an array of sensors receives the signal emitted by the source itself. From the received signal, the range-difference measurements $r_{ij}$ which is the difference in the range between the ${i}$th and the ${j}$th sensor i.e.,  $r_{ij} = r_{i} - r_{j}$ is estimated. Note that, since $r_{ij} = -r_{ji}$, the number of distinct range-differences obtained from an array of $m$ sensors is $\dfrac{m(m-1)}{2}$. However, the estimated range and range-difference measurements are usually noisy due to the measurement errors. In this paper, we consider the problem of localizing the source using the noisy range and range-difference measurements with the assumption that the measurements are obtained in a line-of-sight scenario with known sensor coordinates.    \\
\\
Each range and range-difference measurement geometrically correspond to a circle and a hyperbola, respectively and hence conventionally, the position of the source was found by the method of trilateration. However, this method puts a restriction on the minimum number of sensors required to localize the source uniquely \cite{redundant}. Alternatively, with the assumption that the noise in the range or range-difference measurements follow Gaussian distribution, the least-squares method is adopted to estimate the source coordinates. The resulting problems are non-convex and non-smooth and therefore finding a closed-form solution is difficult. To make the problem differentiable, the measurements are usually ``squared''  and then the source position is estimated in the least-squares sense using the squared measurements \cite{SFP_SWLS, SI, projection, foy, lc, beck}.  Although squaring the measurements simplifies the problem, the source position estimated using this method is influenced more by large errors \cite{myths}. Only some of the methods, such as the methods in \cite{SDR, admm, worst_case, kalman_filter} localize the source without squaring the measurements. Moreover, in the case of source localization using the range-difference measurements, the above mentioned methods localize the source by choosing one of the sensors as a reference sensor; which can affect the positioning accuracy \cite{projection}. This could be because by choosing a reference sensor only $(m-1)$ range-difference measurements are used while a reference-free method uses $\dfrac{m(m-1)}{2}$ range-difference measurements. Only a few methods, such as the methods proposed in  \cite{refreeold}, \cite{amar} and \cite{ono1}, localize the source without choosing a reference sensor. We discuss in detail the existing methods which localize the source using the range and range-difference measurements in section. \ref{sec_range} and \ref{sec_range_difference}, respectively.   \\
\\
In this paper, we estimate the source position iteratively without squaring the noisy range or range-difference measurements and without choosing a reference sensor using Majorization Minimization (MM) approach \cite{tutorial}, \cite{mm}. The significant contributions of this paper are: 
\begin{enumerate}
\item{We propose an algorithm \textbf{So}urce \textbf{L}ocalization \textbf{V}ia an \textbf{It}erative technique (\textbf{SOLVIT}) to localize the source using range-difference measurements without choosing a reference sensor by MM approach.}
\item{The monotonicity and convergence to the stationary point are proved for \textbf{SOLVIT}.}
\item{The Cramer-Rao lower bound for source localization from range-difference measurements is derived.}
\item{\textbf{SOLVIT} is compared with the existing methods via various computer simulations and in a real-life scenario.}
\item{We also localize the source using range measurements and rederive the Standard Fixed Point (SFP) algorithm developed in \cite{SFP_SWLS} using the Majorization Minimization technique and show a less intricate proof of convergence.}
\end{enumerate}

The paper is organized as follows. We first formulate and discuss the related literature on the problem of source localization from noisy range and range-difference measurements in section. \ref{sec_range} and section. \ref{sec_range_difference}, respectively. Next, in section. \ref{sec:2}, we briefly introduce Majorization Minimization. In section. \ref{sec:3}, we present our main contribution - \textbf{SOLVIT}; an iterative method to localize the source using the noisy range-difference measurements. We discuss its complexity followed by which we show how to initialize the proposed algorithm. We also discuss the convergence of the algorithm. At the end of  section. \ref{sec:3}, we prove that the SFP algorithm falls under MM and prove its convergence. In section. \ref{sec:4}, we derive the Cramer-Rao Lower Bound and next evaluate the proposed algorithm by simulations and experiments and finally conclude the paper in section. \ref{sec:5}.\\
\\
Throughout the paper, {\bf{bold}} capital and {\bf{bold}} small letter denote matrix and vector, respectively. A scalar is denoted by a small letter. The value taken by ${\bx}$ at the ${k^{th}}$ iteration is denoted by ${\bx^{k}}$. Euclidean  and Frobenius norm is denoted by ${\| . \|}_{2}$ and $\|.\|_{F}$, respectively. 
\section{Source Localization from Range Measurements}\label{sec_range}
\subsection{Problem Formulation}
In the case of source localization using range-based measurements, an array of ${m}$ sensors transmits a signal towards the object of interest, and each sensor receives the signal reflected by the object. Then, the Time of Arrival (TOA) of the source signal at the ${i}$th sensor is estimated by cross-correlating the ${i}$th transmitted and received signals. Assuming constant velocity propagation medium, range ${r_{i}}$ can be estimated at each sensor. However, due to measurement errors, each range ${r_{i}}$ is only approximately estimated. Specifically, let  $\{\by_{i} \in \mathbb{R}^{n \times 1}\}$ denote the known coordinate of the ${i}$th sensor and $\{\bx \in \mathbb{R}^{n \times 1}\}$ be the unknown source coordinate vector. Then, the approximate or noisy range measurement ${r_{i}}$ is given by:
\begin{equation}\label{p1}
\begin{array}{ll}
r_{i} = {\|\bx - \by_{i}\|}_{2} +\varepsilon_{i}, \quad   i =1,2,\cdots m
\end{array}
\end{equation}
where $\boldsymbol{\varepsilon}=[\varepsilon_{1},\varepsilon_{2}\cdots\varepsilon_{m}]^{T}$ is the error vector. Hence, in range based source localization, the problem is to estimate the source position given the noisy range measurements. A straightforward approach to estimate the source position is by minimizing the sum of the squared errors using the least-squares approach. Mathematically, this can be written as:
\begin{equation} \label{eq:11}
\begin{array}{ll}
\textrm{R-LS:} \quad \underset{\bx}{\rm minimize} \: \{f_{_{\rm R-LS}}\left(\bx\right) \overset{\Delta} = \displaystyle \sum_{i=1}^{m}  \left(r_{i} - {{\|{\bx -\by_{i}}}\|}_{2}\right)^{2}\}
\end{array}
\end{equation}
When $\boldsymbol{\varepsilon}$ follows Gaussian distribution and the errors $[\varepsilon_{i}]_{i=1}^{m}$ are independent of each other, then this approach also corresponds to the maximum likelihood estimation (MLE) problem of estimating ${\bx}$ from (\ref{eq:11}). On expanding $f_{_{\rm R-LS}}\left(\bx\right)$ we get: 
\begin{equation} \label{eq:expand}
\begin{array}{ll}
f_{_{\rm R-LS}}\left(\bx\right) =\displaystyle \sum_{i=1}^{m}  \left(r_{i}^{2} - 2{r_{i}\|\bx-\by_{i}\|}_{2} +{\|\bx-\by_{i}\|}_{2}^{2}\right)
\end{array}
\end{equation}
Note that the above objective function has $-{\|\bx-\by_{i}\|}_{2}$ which is both non-smooth and non-convex function, which makes the R-LS problem challenging to solve. 
\subsection{Related Work}
Geometrically, for ${n = 2}$, each range measurement defines a circle, and the solution for the R-LS problem lies at the intersection of the circles, for which one must have at least \emph{three} sensors \cite{redundant}. The authors in \cite{SDR} tried to solve the R-LS problem numerically by constructing a semidefinite relaxation (SDR) of the R-LS problem. Nevertheless, since SDR is a relaxation of the original problem, it does not always guarantee an optimal solution. Luke et al. \cite{admm} solved (\ref{eq:11}) using alternating directions method of multipliers (ADMM); however, the method suffers from slow convergence \cite{admm_slow}. Beck et al. \cite{SFP_SWLS}, inspired by the famous ``Weiszfeld algorithm'' proposed the - ``Standard Fixed Point algorithm'' to solve the range-based source localization problem. Similar to the Weiszfeld algorithm, it is a fixed point method. Recently, Tzoreff et al. \cite{em} used the Expectation-Maximization algorithm to estimate the source position directly from the range measurements. An alternate approach to localize the source is by solving the following least-squares problem using the squared range measurements: \\
\begin{equation} \label{eq:11_square}
\begin{array}{ll}
\textrm{SR-LS:} \quad \underset{\bx}{\rm minimize} \: \displaystyle \sum_{i=1}^{m}  \left(r_{i}^{2} - {{\|{\bx -\by_{i}}}\|}_{2}^{2}\right)^{2}
\end{array}
\end{equation}
The above problem is differentiable and is hence easier to solve  when compared to the R-LS problem. However, the solution obtained by solving the above problem is less accurate when compared to the solution obtained by solving the R-LS problem \cite{SFP_SWLS}. Also, the solution obtained by solving the SR-LS problem does not correspond to the Maximum Likelihood estimator. Hence, in this paper, we consider the R-LS problem and rederive the SFP algorithm using the MM framework. By doing so, we are able to show a more straightforward proof of convergence for the SFP algorithm. 
\section{Source Localization from Range-Difference Measurements}\label{sec_range_difference}
\subsection{Problem Formulation}
In the case of source localization using range-difference measurements, the array of ${m}$ sensors only receives the source signal. The signal received at each sensor is the delayed source signal itself. The \emph{Time Difference of Arrival} (TDOA) of the received signal at the ${i}$th and the ${j}$th sensors can be estimated by cross-correlating them. Assuming constant velocity of propagation, from the TDOA, range-difference ${r_{ij}}$ can be estimated. Similar to the R-LS problem, the range-differences are only approximately estimated, i.e. 
\begin{equation}
\begin{array}{ll} 
r_{ij} ={ \|\bx - \by_{i}\|}_{2}+\varepsilon_{i}-{ \|\bx - \by_{j}\|}_{2}-\varepsilon_{j}\quad   i, j=1,2,\cdots m\\
\\
\quad\,\, ={ \|\bx - \by_{i}\|}_{2}-{ \|\bx - \by_{j}\|}_{2}+ \varepsilon_{ij} \quad\quad\quad   i, j=1,2,\cdots m
\end{array}
\end{equation}
where $\varepsilon_{ij}=\varepsilon_{i}-\varepsilon_{j}$ is the range-difference noise and is assumed to be Gaussian distributed. Hence in range-difference based source localization, the problem is to estimate the source position given the noisy range-differences. Also, note that :
\begin{equation}
\begin{array}{ll}
r_{ij}=-\left({ \|\bx - \by_{j}\|}_{2}-{ \|\bx - \by_{i}\|}_{2} +\varepsilon_{ji}\right)=-r_{ji}
\end{array}
\end{equation}
Hence, for $m$ sensors, the number of possible distinct or positive noisy range-differences is $\dfrac{m(m-1)}{2}$.  As in the range based source localization, the source position can be estimated from the noisy range-differences by the least-squares criteria. Mathematically, the problem is written as: 
\begin{equation} \label{eq:12}
\begin{array}{ll}
\textrm{RD-LS:} \quad \underset{\bx}{\rm minimize}\{f_{_{\rm RD-LS}}\left(\bx\right) \overset{\Delta} = \\
\displaystyle\sum_{\substack{i, j = 1 \\ r_{ij}>0}}^{m}  \left(r_{ij} - \left({\|\bx -\by_{i}\|}_{2} - \|{\bx -\by_{j}\|}_{2}\right)\right)^{2}\}
\end{array}
\end{equation}
Note that in the above problem, we include only the distinct range-differences, i.e. from the $m \times m$ range-differences only the $\dfrac{m(m-1)}{2}$ positive range-differences are chosen and minimized in (\ref{eq:12}). On expanding $f_{_{\rm RD-LS}}\left(\bx\right)$ we get:
\begin{equation}
\begin{array}{ll}
f_{_{\rm RD-LS}}\left(\bx\right) = \displaystyle{\sum_{\substack{i, j = 1 \\ r_{ij}>0}}^{m}} \bigg(r_{ij}^{2} +{\|\bx-\by_{i}\|}_{2}^{2}+{\|\bx-\by_{j}\|}_{2}^{2}\\-2r_{ij}{\|\bx-\by_{i}\|}_{2}
+2r_{ij}{\|\bx-\by_{j}\|}_{2}-2{\|\bx-\by_{i}\|}_{2}{\|\bx-\by_{j}\|}_{2}\bigg)
\end{array}
\end{equation}
Similar to the R-LS problem, the above cost function has ${-{\|\bx-\by_{i}\|}_{2}}$, which is both non-convex and non-smooth function.  Moreover, $-{\|\bx-\by_{i}\|}_{2}{\|\bx-\by_{j}\|}_{2}$ term makes it even more challenging to solve the RD-LS problem when compared to the R-LS problem.
\subsection{Related Work}
Geometrically, each range-difference equation corresponds to a hyperbola (for ${n = 2}$), and hence this problem is also called \emph{hyperbolic positioning} problem \cite{redundant} and to solve the problem uniquely, one requires \emph{four} sensors \cite{redundant}. Numerically, the state-of-the-art algorithms solve the problem in (\ref{eq:12}) by squaring the range-difference measurements and choosing one of the sensors as a reference sensor. Assuming, that the first sensor is the reference sensor located at the origin, then the range-difference ${r_{i1}}$ between the ${i}$th sensor and the sensor at origin is given by:
\begin{equation}
r_{i1} ={\|\bx -\by_{i}\|}_{2}- {\|\bx\|}_{2},   \quad i = 2,\cdots m
\end{equation}
which after squaring and re-arranging:
\begin{equation}\label{square}
-2\by_{i}^{T}\bx -2r_{i1}{\|\bx\|}_{2} = r_{i1}^{2} -{\|\by_{i}\|}_{2}^{2},  \quad i = 2,\cdots m
\end{equation}
The above relation is approximate due to measurement errors. Hence, in the reference-based squared range-differences, the source is localized by the minimization of the following least-squares criterion:
\begin{equation} \label{eq:13}
\begin{array}{ll}
\textrm{SRD-LS:} \quad \underset{\bx}{\rm minimize} \:\displaystyle\sum_{i=2}^{m}  \left(-2{\by_{i}}^{T}\bx -2r_{i1}{\|\bx\|}_{2}-g_{i}\right)^{2}
\end{array}
\end{equation}
where ${g_{i} = r_{i1}-{\|y_{i}\|}_{2}^{2}}$. By squaring the range-differences and then minimizing the squared errors, the above problem smartly eliminates the product of norm terms, which was present in the RD-LS problem. Hence, the SRD-LS problem is easier to solve when compared to the RD-LS problem (\ref{eq:12}). Many methods have been proposed to solve the SRD-LS problem. Smith et al. \cite{SI} reformulated the non-convex problem in (\ref{eq:13}) into a convex problem with two unknowns, ${\bx}$ and ${{\|\bx\|}_{2}}$. Then, by two-stage unconstrained minimization, ${\bx}$ was estimated. This method is called the \emph{spherical interpolation} method. Friedlander et al. \cite{projection} also solved the problem similar to Smith et al.; however, they eliminated ${{\|\bx\|}_{2}}$ by using the orthogonal projection matrix and then estimated the source position using least squares. Linear correction method \cite{lc} and the method proposed by Beck et al. \cite{beck} converted the SRD-LS problem into a constrained optimization problem and solved it by the bisection algorithm. Another approach is to linearize the SRD-LS problem using Taylor Series \cite{foy} and solve the problem iteratively. Shi et al. \cite{worst_case} localized the source by minimizing the worst-case positioning error. They solved the problem by relaxing the non-convex problem into a convex problem by semidefinite relaxation. Recently, based on extended Kalman Filter, Tovkach et al. \cite{kalman_filter} developed a recurrent algorithm to localize the source. \\
\\
All these methods estimate the position using a reference sensor. However, in \cite{projection}, it was found that choosing different reference sensors can affect the positioning accuracy. Only a few methods localize the source without choosing a reference sensor. Schmidt in \cite{refreeold} proposed a two-step reference-free source localization method. In the first step, the noisy TDOA measurements are denoised in the least-squares sense. Then, in the second step, using the denoised TDOA, the source is localized using Location on the Conic Axis (LOCA) algorithm. Instead of the traditional hyperbolic intersection principle, LOCA localizes the source from the intersection of lines (which corresponds to the conic axis) obtained from each set of three sensors. Ono et al. in \cite{ono1} localized the source iteratively using auxiliary approach without squaring the range-difference measurements and without choosing a reference sensor. Amar et al. \cite{amar} estimated the source position without choosing the reference sensor using the squared range-differences in the least-squares criterion. Nevertheless, the source position estimate obtained by solving the SRD-LS problem is influenced more by large errors \cite{myths} and is sub-optimal in the Maximum Likelihood sense as the errors in (\ref{square}) are not independent \cite{mle_suboptimal}. In this paper, we show a novel way to tackle the RD-LS problem and hence, estimate the source position without choosing a reference sensor and without squaring the range-difference measurements. \\
\section{Majorization Minimization}\label{sec:2}
Majorization Minimization (MM) is an iterative procedure that is mostly used to solve a non-convex, non-smooth or even a convex problem more efficiently. In MM, instead of minimizing the difficult optimization problem $f(\bx)$ directly, a ``surrogate'' function, which  majorizes the problem (at a given value of $\bx$ = $\bx^{k}$) is minimized at every iteration. The surrogate function $g(\bx|\bx^{k})$ is the global upper bound of the objective function $f(\bx)$ i.e., it satisfies the following properties:
\begin{equation}  \label{eq:21}
g\left(\bx^{k}|\bx^{k}\right) = f\left(\bx^{k}\right) 
\end{equation}
\begin{equation}\label{eq:22}
g\left(\bx|\bx^{k}\right) \geq f\left(\bx\right) 
\end{equation}
where, ${\bx^{k}}$ is the value taken by $\bx$ at the $k^{th}$ iteration. Hence, the MM algorithm generates a sequence of points ${\{\bx^{k}\}}$ according to the following rule: 
\begin{equation}  \label{eq:23}
\bx^{k+1} \in \underset{\bx}{\rm arg\:min} \: g\left(\bx|\bx^{k}\right)
\end{equation}
By using (\ref{eq:21}), (\ref{eq:22}) and (\ref{eq:23}), it can be proved that MM monotonically decreases the objective function \cite{mm}. The convergence rate and computational complexity of the algorithm depend on how well one formulates the surrogate function. To have lower computational complexity, the surrogate function must be easy to minimize. However, the convergence rate depends on how well the surrogate function follows the shape of the objective function $f(\bx)$. Hence, the novelty of the MM algorithm lies in the design of the surrogate function. To design a surrogate function, there are no set steps to follow. However, there are a few papers that give guidelines for designing various surrogate functions \cite{tutorial}, \cite{mm}.

\section{Algorithms for source localization problems based on MM}\label{sec:3}
In this section, we first propose a novel algorithm, \textbf{SOLVIT}, for the RD-LS  problem based on MM. We then discuss its convergence and show that it converges to the stationary point of the problem in (\ref{eq:12}). At the end of this section, we solve the R-LS problem and rederive the SFP algorithm using MM and prove its convergence for the sake of completeness. 
\subsection{\textbf{SOLVIT}: \textbf{So}urce \textbf{L}ocalization \textbf{v}ia an \textbf{It}erative technique}
We now propose \textbf{SOLVIT}, an iterative algorithm to solve the RD-LS problem based on MM. On expanding $f_{_{\rm RD-LS}}\left(\bx\right)$, in (\ref{eq:12}) we get: 
\begin{equation}\label{expand}
\begin{array}{ll}
f_{_{\rm RD-LS}}\left(\bx\right) = \displaystyle{\sum_{\substack{i, j = 1 \\ r_{ij}>0}}^{m}} \bigg(r_{ij}^{2} +{\|\bx-\by_{i}\|}_{2}^{2}+{\|\bx-\by_{j}\|}_{2}^{2}\\-2r_{ij}{\|\bx-\by_{i}\|}_{2}
+2r_{ij}{\|\bx-\by_{j}\|}_{2}-2{\|\bx-\by_{i}\|}_{2}{\|\bx-\by_{j}\|}_{2}\bigg)
\end{array}
\end{equation}
The norm and the product of norm terms in $f_{_{\rm RD-LS}}\left(\bx\right)$ are not continuously differentiable, and hence, getting a closed-form solution for the RD-LS problem is difficult from the first-order conditions. Hence, we form a surrogate function $g\left(\bx|\bx^{k}\right)$ to majorize these non-differentiable terms and hence majorize$f_{_{\rm RD-LS}}\left(\bx\right)$ based on the following lemmas: 
\\
\begin{lemma} \label{lemma 1}
Given any ${\bx = \bx^{k}}$, $-{\|\bx-\by_{i}\|}_{2}$  can be upper  bounded as 
\begin{equation}\label{eq:31}
\begin{array}{ll}
  - {\|\bx-\by_{i}\|}_{2}  \leq - \dfrac{{\left(\bx-\by_{i}\right)^{T}}\left(\bx^{k}-\by_{i}\right)}{{\|\bx^{k}-\by_{i}\|}_{2}}
\end{array}
\end{equation}
The upper bound for $ - {\|\bx-\by_{i}\|}_{2}$ is linear and differentiable in $\bx$.
\end{lemma}
\begin{IEEEproof}
\cite{mm} By Cauchy-Schwarz  inequality we have:
\begin{equation*}
{\left(\bx-\by_{i}\right)^{T}}{\left(\bx^{k}-\by_{i}\right)} \leq {\|\bx-\by_{i}\|}_{2}{\|\bx^{k}-\by_{i}\|}_{2}
\end{equation*}
After rearranging we get:
\begin{equation*} 
{\|\bx-\by_{i}\|}_{2} \geq \dfrac{{\left(\bx-\by_{i}\right)^{T}}\left(\bx^{k}-\by_{i}\right)}{{\|\bx^{k}-\by_{i}\|}_{2}}
\end{equation*}
By taking the negative sign into account, we get the inequality. 
\end{IEEEproof}
\begin{lemma} \label{lemma 2}
Given any ${\bx = \bx^{k}}$, ${\|\bx-\by_{j}\|}_{2}$  can be upper  bounded as 
\begin{equation}\label{eq:41}
\begin{array}{ll}
 {\|\bx-\by_{j}\|}_{2} \leq  {\|\bx^{k}-\by_{j}\|}_{2} + \dfrac{{\|\bx-\by_{j}\|}_{2}^{2}-{\|\bx^{k}-\by_{j}\|}_{2}^{2}}{2{\|\bx^{k}-\by_{j}\|}_{2}}
\end{array}
\end{equation}
The upper bound for ${\|\bx-\by_{j}\|}_{2}$ is quadratic and differentiable in $\bx$.
\end{lemma}
\begin{IEEEproof}
Let $t  ={\|\bx-\by_{j}\|}_{2}^2$. Therefore,
\begin{equation*}
{\|\bx-\by_{j}\|}_{2} = \sqrt{t}    
\end{equation*}
Since the square root  is concave in $\mathcal{R}$ \cite{boyd}, linearizing it around ${t^{k}}$ gives the following inequality
\begin{equation*}
\sqrt{t} \leq \sqrt{t^{k}} + \dfrac{t - t^{k}}{2\sqrt{t^{k}}}
\end{equation*}
Substituting back for ${t}$, the inequality can be attained. 
\end{IEEEproof}
\begin{lemma} \label{lemma 3}
Given any ${\bx = \bx^{k}}$, ${-{\|\bx - \by_{i}\|}_{2}{\|\bx-\by_{j}\|}_{2}} $ can be upper bounded as
\begin{equation}\label{eqn:42} 
\begin{array}{ll}
-{\|\bx-\by_{i}\|}_{2}{\|\bx-\by_{j}\|}_{2} \leq - \left(\bx-\by_{j}\right)^{T}\bQ {\left(\bx-\by_{i}\right)}
\end{array}
\end{equation}
where
\begin{equation}\label{eqn:43} 
\begin{array}{ll}
\bQ = \dfrac{\left(\bx^{k}-\by_{j}\right)\left(\bx^{k}-\by_{i}\right)^{T}}{{\|\bx^{k} - \by_{j}\|}_{2}{\|\bx^{k}- \by_{i}\|}_{2}}
\end{array}
\end{equation}
The upper bound is quadratic and differentiable in ${\bx}$.
\end{lemma}
\begin{IEEEproof}
The inequality in (\ref{eqn:42}) can be proved by solving the following minimization problem:
\begin{equation} 
\begin{array}{ll}
\underset{\{\bQ,\,\|\bQ\|_{F}=1\}}{\rm minimize}\: \| \bQ- {\left(\bx-\by_{j}\right)}\left(\bx- \by_{i}\right)^{T}\|^{2}_{F}
\end{array}
\end{equation}
After removing the terms independent of ${\bQ}$ and by exploiting the constraint, the above problem can be re-written as:
\begin{equation}
\begin{array}{ll}
\underset{\{\bQ,\,\|\bQ\|_{F}=1\}}{\rm maximize}\: \textrm{Tr}\left(\bQ\left(\left(\bx- \by_{j}\right)\left(\bx-\by_{i}\right)^{T}\right)^{T}\right)
\end{array}
\end{equation}
which can be further re-written as:
\begin{equation}
\begin{array}{ll}
\underset{\{\bQ,\,\|\bQ\|_{F}=1\}}{\rm maximize}\: \left(\bx-\by_{j}\right)^{T}\bQ \left(\bx- \by_{i}\right)
\end{array}
\end{equation}
The above problem can be solved using the method of Lagrange multipliers and its solution is given by:
\begin{equation} 
\begin{aligned}
\bQ^{*} = \dfrac{(\bx-\by_{j})(\bx-\by_{i})^{T}}{{\|\bx - \by_{j}\|}_{2}{\|\bx- \by_{i}\|}_{2}}
 \end{aligned}
\end{equation}
and the value the objective function takes at the maximum is ${{\|\bx-\by_{j}\|}_{2}{\|\bx-\by_{i}\|}_{2}}$. Therefore, the maximum value $\left(\bx-\by_{j}\right)^{T}\bQ \left(\bx- \by_{i}\right)$, for $\bQ$ equal to $\bQ^{*}$, is ${{\|\bx-\by_{j}\|}_{2}{\|\bx-\by_{i}\|}_{2}}$ and hence the inequality. 
\end{IEEEproof}
Using lemma \ref{lemma 1}, lemma \ref{lemma 2} and lemma \ref{lemma 3} in (\ref{expand}) we get the following surrogate function:
\begin{equation}\label{eqn:45}
\begin{array}{ll} 
g\left(\bx|\bx^{k}\right)=\displaystyle{\sum_{\substack{i, j = 1 \\ r_{ij}>0}}^{m}} \bigg(r_{ij}^{2}+{\|\bx-\by_{i}\|}_{2}^{2} + {\|\bx-\by_{j}\|}_{2}^{2} - 2r_{ij}\bw_{i}^{T}\\\left(\bx-\by_{i}\right)  
+s_{ij}{\|\bx- \by_{j}\|}_{2}^{2} - 2\left((\bx-\by_{j})^{T}\bQ_{ij}(\bx-\by_{i})\right)\bigg)
\end{array}
\end{equation}
where
\begin{equation} \label{eqn:46} 
\begin{array}{ll}
\bw_{i} = \dfrac{\bx^{k}- \by_{i}}{{\|\bx^{k}- \by_{i}\|}_{2} }
\end{array}
\end{equation}
\begin{equation} \label{eqn:47} 
\begin{array}{ll}
s_{ij} = \dfrac{r_{ij}}{{\|\bx^{k}- \by_{j}\|}_{2}}
\end{array}
\end{equation}
\begin{equation} \label{eqn:48} 
\begin{array}{ll}
\textbf{Q}_{ij}=\dfrac{\left(\bx^{k}- \by_{j}\right)\left(\bx^{k}- \by_{i}\right)^{T}}{{\|\bx^{k}- \by_{j}\|}_{2}{\|\bx^{k} - \by_{i}\|}_{2}}
\end{array}
\end{equation}
Note that $g\left(\bx|\bx^{k}\right) \geq f_{_{\rm RD-LS}}\left(\bx\right)$ for all $\bx$, and equality is achieved at $\bx = \bx^{k}$. The surrogate function $g\left(\bx|\bx^{k}\right)$ in (\ref{eqn:45}) is differentiable in ${\bx}$. Hence at any iteration, given $\bx^{k}$, the surrogate minimization problem for (\ref{eq:11}) becomes:
\begin{equation}
\begin{array}{ll}
\underset{\bx}{\rm minimize} \: g\left(\bx|\bx^{k}\right)
\end{array}
\end{equation}
where $g\left(\bx|\bx^{k}\right)$ is given by (\ref{eqn:45}). The above problem in contrast to the RD-LS problem admits a closed-form solution and is given by:
\begin{equation} \label{eqn:49} 
\begin{array}{ll}
\bx^{k+1}\overset{\Delta} = \bA\bb = \left({\displaystyle\sum_{\substack{i, j = 1 \\ r_{ij}>0}}^{m}} \bM_{ij}\right)^{-1}\left({\displaystyle\sum_{\substack{i, j = 1 \\ r_{ij}>0}}^{m}}\bp_{ij}\right) 
\end{array}
\end{equation}
where
\begin{equation}\label{eqn:50}  
\begin{array}{ll}
\bM_{ij} =   \left(2\textbf{I} +s_{ij}\textbf{I} - \left(\bQ_{ij}+ \bQ_{ij}^{T}\right)\right)\\
\end{array}
\end{equation}
\begin{equation} \label{eqn:51} 
\begin{array}{ll}
\bp_{ij} =   (\by_{i} + \by_{j} + r_{ij}\bw_{ij} + s_{ij}\by_{j} - \bQ_{ij}\by_{i} - \bQ_{ij}^{T}\by_{j})
\end{array}
\end{equation}
For $\bA$ to exist always, it is sufficient to prove that ${\bM_{ij}}$ is positive definite for values of ${i, j}$ $\in$ $\{1, 2\cdots m\}$ and ${r_{ij}} >0$. ${\bM_{ij}}$ is positive definite if the maximum eigen value of $(\bQ_{ij}+ \bQ_{ij}^{T})$ is less than $(2 +s_{ij})$. The same is proved in the following lemma.

\begin{lemma}
$\lambda_{max}$ of $(\bQ_{ij}+ \bQ_{ij}^{T})$ is less than $(2 +s_{ij})$ $\forall$ $\{i, j\}$ $\in$ $\{1, 2\cdots m\}$ and ${r_{ij}} >0$
\end{lemma}
\begin{IEEEproof}
$\bQ_{ij}$ matrix can be written as ${\bu_{j}\bv_{i}^{T}}$  where 
\begin{equation}\label{u1}
\begin{array}{ll}
\bu_{j}= \dfrac{(\bx^{k}- \by_{j})}{{\|\bx^{k} - \by_{j}\|}_{2}}
\end{array}
\end{equation}
\begin{equation}\label{u2}
\begin{array}{ll}
\bv_{i} =\dfrac{(\bx^{k}- \by_{i})}{{\|\bx^{k} - \by_{i}\|}_{2}}
\end{array}
\end{equation}
The maximum eigen-value of $(\bQ_{ij}+ \bQ_{ij}^{T})$, i.e., $\left(\bu_{j}\bv_{i}^{T} + \bv_{i}\bu_{j}^{T}\right)$ can be obtained by solving the following optimization problem: 
\begin{equation}\label{withoutnorm}
\begin{array}{ll}
\underset{\|\bz\|=1}{\rm\, max} \: \bz^{T}\left(\bu_{j}\bv_{i}^{T} + \bv_{i}\bu_{j}^{T}\right)\bz
\end{array}
\end{equation}
where the maximizer would be the eigen-vector corresponding to the maximum eigen-value of $\left(\bu_{j}\bv_{i}^{T} + \bv_{i}\bu_{j}^{T}\right)$. Now, consider the following: 
\begin{equation}\label{withnorm}
\begin{array}{ll}
\bz^{T}\left(\bu_{j}\bv_{i}^{T} + \bv_{i}\bu_{j}^{T}\right)\bz = |\bz^{T}\left(\bu_{j}\bv_{i}^{T} + \bv_{i}\bu_{j}^{T}\right)\bz|\\
\quad \quad \overset{(a)}{\leq}  2|\bz^{T}\left(\bu_{j}\bv_{i}^{T}\right)\bz|\overset{(b)}{\leq} 2\|\bu_{j}\|\|\bg\|\|\bv_{i}\|\|\bg\|\overset{(c)}{\leq} 2. 
\end{array}
\end{equation}
where the inequality (a) is by Triangle law of inequality, and the next inequality (b) is by Cauchy-Schwartz. The inequality in (c) is by exploiting the fact that $\bz$, $\bu$ and $\bv$ are unit vectors.
Hence, the maximum eigen-value of $\left(\bQ_{ij}+ \bQ_{ij}^{T}\right)$ is always less than or equal to two, which is less than two plus ${s_{ij}}$ (which is always positive).
\end{IEEEproof}
\subsection{Proof of Convergence}\label{sec:5}
Given that the proposed algorithm is based on the Majorization Minimization framework, it is ensured that the sequence of points $\{\bx^{k}\}$ monotonically decreases the {RD-LS} problem. Moreover, the {RD-LS} problem is bounded below by zero. Hence, the sequence $f_{_{\rm RD-LS}}\left(\bx\right)$ generated by the proposed algorithm will, at the least, converge to a finite value.\\

We now show that the sequence $\{\bx^{k}\}$ will converge to the stationary point.  A point ${\bx}$ is called stationary if:
\begin{equation}
\begin{array}{ll}
f_{_{\rm RD-LS}}'(\bx;\bd) \geq 0
\end{array}
\end{equation}
where $f_{_{\rm RD-LS}}'(\bx;\bd)$ is the directional derivative of function $f_{_{\rm RD-LS}}\left(\cdot\right)$ at point $\bx$ in the direction $\bd$ and is defined as:
\begin{equation}
\begin{array}{ll}
f_{_{\rm RD-LS}}'(\bx;\bd) =\underset{\lambda \rightarrow 0}{\lim} \: \textrm{inf}\dfrac{f_{_{\rm RD-LS}}(\bx+\lambda\bd) - f_{_{\rm RD-LS}}(\bx)}{\lambda}
\end{array}
\end{equation}
From the monotonic property of MM, we have:
\begin{equation}\label{conv1}
\begin{array}{ll}
f_{_{\rm RD-LS}}\left(\bx^{0}\right) \geq f_{_{\rm RD-LS}}\left(\bx^{1}\right) \geq f_{_{\rm RD-LS}}\left(\bx^{2}\right) \cdots
\end{array}
\end{equation}
Assume a subsequence ${\bx^{r_{j}}}$ converging to a limit point $\bz$. Then, from (\ref{eq:21}), (\ref{eq:22}) and from (\ref{conv1}) we obtain:
\begin{equation}
\begin{array}{ll}
g\left(\bx^{r_{j+1}}|\bx^{r_{j+1}}\right)= f_{_{\rm RD-LS}}\left(\bx^{r_{j+1}}\right) \leq f_{_{\rm RD-LS}}\left(\bx^{r_{j}+1}\right) \leq \\
g\left(\bx^{r_{j}+1}|\bx^{r_{j}}\right)\leq g\left(\bx|\bx^{r_{j}}\right)
\end{array}
\end{equation}
which implies $g'(\bz|\bz) \geq 0$ and $f'(\bz) \geq 0$ \cite{convergence}. 
Hence, $\bz$ is the stationary point of $f_{_{\rm RD-LS}}\left(\cdot\right)$ and therefore the proposed algorithm converges to the stationary point of $f_{_{\rm RD-LS}}\left(\bx\right)$.
\subsection{Initialization of \textbf{SOLVIT}}\label{initialize}
Proper initialization of \textbf{SOLVIT} is crucial for the algorithm to converge to the stationary point. This is because $f_{_{\rm RD-LS}}\left(\bx\right)$ in (\ref{eq:12}) becomes flat as its argument goes to infinity. Hence, \textbf{SOLVIT} should never be initialized in such a region. We now discuss a way to initialize the \textbf{SOLVIT} algorithm. The main aim is to initialize \textbf{SOLVIT} possibly closer to the true solution and to avoid the flatter regions of $f_{_{\rm RD-LS}}\left(\bx\right)$. This can be done by exploiting the fact that each range-difference equation corresponds to a hyperbola (for $n=2$), and the true source position $\bx = (x,y)$ is a point of intersection of the hyperbolas - which means $\bx$ lies on all the $^mC_2$ hyperbolas. Hence, \textbf{SOLVIT} can be initialized anywhere on a randomly chosen hyperbola, which makes it closer to the true solution. Therefore to get a good initial point, we do grid search over small values $x$ and $y$ (so that \textbf{SOLVIT} does not get stuck  in the flat-region of the objective function) which lie on the randomly chosen hyperbola and select the $\bx^{int}$ with minimum value for the function $f_{_{\rm RD-LS}}\left(\bx\right)$ in (6).\\
\subsection{Computational Complexity of \textbf{SOLVIT}}\label{sec:6}
Let $\hat{m}$ be the number of positive range - differences and is equal to $\dfrac{m(m-1)}{2}$. 
The complexity in calculating $\bA$ is equal to $\mathcal{O}\left(n^{3}\right)$. The complexity of implementing  $\bw_{i}$ from  (\ref{eqn:46}) is $\mathcal{O}(n\hat{m})$, $s_{ij}$ from (\ref{eqn:47}) is $\mathcal{O}(n\hat{m})$, and $\bQ_{ij}$ for specific $i$, $j$ from (\ref{eqn:48}) is ${\mathcal{O}\left(n^{2}\right)}$. Then, the complexity of calculating summation of $\bM_{ij}$ for $\{i,j =1,2,\cdots\hat{m}\}$ is $\mathcal{O}(n^2\hat{m})$. Taking into account the complexity of matrix-vector multiplication, the complexity of implementing \textbf{SOLVIT} is $\mathcal{O}\left(n^{2}\hat{m}\right)$.\\
We now discuss the complexity of finding a good initial point, as proposed in the previous section. The complexity of finding a point $\bx^{int} = [x, y]$ lying on the randomly chosen hyperbola is $\mathcal{O}(n)$. Then the cost of evaluating the objective function in (6) for this $\bx^{int}$ is $\mathcal{O}(\hat{m}n)$. Assuming that the size of the grid search is $l$, then including the cost of finding the minimum value, the total computational complexity of finding a good initial point is $\mathcal{O}(l\hat{m}n)$.
\\
\noindent 	The Pseudocode of the proposed algorithm is shown below:
\begin{center}
\begin{tabular}{@{}p{8cm}}
\hline
\hline
\bf{\emph{Pseudocode of \textbf{SOLVIT}}} \\
\hline
\hline
{\bf{Input}}: Sensor locations $(\by_{1},\by_{2}\cdots\by_{m})$ , noisy distinct range differences (${r_{21},r_{31}\cdots r_{\hat{m}\,\hat{m}-1}}$); \\
{\bf{Initialize}}: Set \emph{k} = 0. Initialize ${\bx^{0}}$. \\
{\bf{Repeat}}: \\ 
    for ${i, j = 1:\hat{m}}$\\
    1) Compute $s_{ij}$, $\bw_{i}$, $\bp_{ij}$, $\bu_{j}$ and $\bv_{i}$ and  from (\ref{eqn:47}), (\ref{eqn:48}), (\ref{eqn:51}), (\ref{u1}) and (\ref{u2}) respectively.\\
    end\\
    2) Compute $\bA = \left({\displaystyle\sum_{\substack{i, j = 1 }}^{\hat{m}}} \bM_{ij}\right)^{-1}$ using Woodbury Matrix Identity  and compute $\bb = {\displaystyle\sum_{\substack{i, j = 1}}^{\hat{m}}}\bp_{ij}$\\
    3) $\bx^{k+1} =\bA\bb$, $k \leftarrow k+1$, {\bf{until $\left|\dfrac{f_{_{\rm RD-LS}}({\bx^{k+1}})-f_{_{\rm RD-LS}}({\bx^{k}})}{f_{_{\rm RD-LS}}({\bx^{k}})}\right|<10^{-4}$}}\\
\hline
\hline
\end{tabular}
\end{center}

\subsection{Derivation of Standard Fixed Point (SFP) algorithm using Majorization Minimization}
The SFP algorithm shown in \cite{SFP_SWLS} was developed to localize the source using range measurements. SFP is a fixed point algorithm that was derived using the first-order optimal conditions. Nevertheless, the proof of monotonic convergence of the SFP algorithm is intricate. Here, we rederive the SFP algorithm using the Majorization Minimization procedure and show a simple and less intricate way to prove the convergence of SFP.\\
\\
On expanding $f_{_{\rm R-LS}}(\bx)$ in (\ref{eq:11}) we get:
\begin{equation}\label{eq:32}
\begin{array}{ll}
f_{_{\rm R-LS}}(\bx) = \displaystyle{\sum_{i=1}^{m}} \left({r_{i}}^{2} - 2{r_{i}{\|\bx-\by_{i}\|}_{2}} +{\|\bx- \by_{i}\|}_{2}^{2}\right)
\end{array}
\end{equation} 
Since $ - {\|\bx - \by_{i}\|}_{2}$ is not continuously differentiable, hence solving the R-LS problem is difficult.  Therefore, the following surrogate function is constructed to majorize $ - {\|\bx - \by_{i}\|}_{2}$ term using lemma \ref{lemma 1} 
\begin{equation}\label{eq:33}
\begin{array}{ll}
g\left(\bx|\bx^{k}\right)=\displaystyle {\sum_{i=1}^{m}}\bigg({r_{i}}^{2} - 2{r_{i}\dfrac{{\left(\bx-\by_{i}\right)}^{T}\left(\bx^{k}-\by_{i}\right)}{{{\|\bx^{k} - \by_{i}\|}_{2}}}} \\\quad \quad \quad+{\|\bx- \by_{i}\|}_{2}^{2}\bigg)
\end{array}
\end{equation}
Note that $g\left(\bx|\bx^{k}\right) \geq f_{_{\rm R-LS}}(\bx)$ for all $\bx$ and the equality is achieved at $\bx = \bx^{k}$. The surrogate function is convex and differentiable in ${\bx}$. Hence at any iteration, the surrogate minimization problem for (\ref{eq:11}) becomes:
\begin{equation}
\begin{array}{ll}
\underset{\bx}{\rm minimize} \: \displaystyle {\sum_{i=1}^{m}} \bigg({r_{i}}^{2} - 2{r_{i}\dfrac{{\left(\bx-\by_{i}\right)}^{T}\left(\bx^{k}-\by_{i}\right)}{{{\|\bx^{k} - \by_{i}\|}_{2}}}}\\\quad \quad \quad+{\|\bx- \by_{i}\|}_{2}^{2}\bigg)
\end{array}
\end{equation}
The above problem has a closed-form minimizer and is given by:
\begin{equation}\label{eq:34}
    \begin{array}{ll}
\bx^{k+1}= \dfrac{1}{m}{\displaystyle\sum_{i=1}^{m}} \left(\by_{i}+r_{i}{\bw}_{i}\right)
\end{array}
\end{equation} 
where
\begin{equation}\label{eq:35}
\begin{array}{ll}
\bw_{i} =\dfrac{\bx^{k}-\by_{i}} {{\|\bx^{k}-{\by_{i}\|}_{2}}}
\end{array}
\end{equation}
The pseudocode of the MM derived SFP algorithm is given:
\begin{center}
\begin{tabular}{ @{}p{8cm} }
\hline
\hline
\bf{Pseudocode of SFP algorithm} \\
\hline
\hline
{\bf{Input}}:  $(\by_{1},\by_{2}\cdots\by_{m})$ ,(${r_{1},r_{2}\cdots r_{m}}$); \\
{\bf{Initialize}}: Set \emph{k} = 0. Initialize ${\bx^{0}}$. \\
{\bf{Repeat}}: \\ 
    for $i = 1:m$\\
    1) Compute ${\bw_{i}}$ from (\ref{eq:35})\\
    2) Compute ${\by_{i} +r_{i}\bw_{i}}$ \\
     end\\
      3) $\bx^{k+1} =\dfrac{1}{m}{\displaystyle\sum_{i=1}^{m}} \left(\by_{i}+r_{i}{\bw}_{i}\right)$,  $k \leftarrow k+1$\\ 
{\bf{until $\left|\dfrac{f_{_{\rm R-LS}}({\bx^{k+1}})-f_{_{\rm R-LS}}({\bx^{k}})}{f_{_{\rm R-LS}}({\bx^{k}})}\right|<10^{-4}$}}\\
\hline
\hline
     \end{tabular}
\end{center}
The update equation for ${\bx}$ in the Standard Fixed Point algorithm found in \cite{SFP_SWLS} is same as (\ref{eq:34}). Hence, the Standard Fixed point algorithm falls under MM. \\
\\
We now prove the convergence of the Standard Fixed Point algorithm. Since we derived the SFP algorithm using the MM framework, it is ensured that the sequence of points $\{\bx^{k}\}$ monotonically decreases the R-LS problem. Similar to the RD-LS problem, the R-LS problem is bounded below by zero. Hence, the sequence $f_{_{\rm R-LS}}(\bx^{k})$ generated by the SFP algorithm will, at the least, converge to a finite value.\\
\\
We now show that the sequence $\{\bx^{k}\}$ will converge to the stationary point. The proof is similar to the convergence proof of the RD-LS problem.  
From the monotonic property of MM, we have:
\begin{equation}\label{conv}
\begin{array}{ll}
f_{_{\rm R-LS}}\left(\bx^{0}\right) \geq f_{_{\rm R-LS}}\left(\bx^{1}\right) \geq f_{_{\rm R-LS}}\left(\bx^{2}\right) \cdots
\end{array}
\end{equation}
Assume a subsequence ${\bx^{r_{j}}}$ converging to a limit point $\bz$. Then, from (\ref{eq:21}), (\ref{eq:22}) and from (\ref{conv}) we obtain:
\begin{equation}
\begin{array}{ll}
g\left(\bx^{r_{j+1}}|\bx^{r_{j+1}}\right)= f_{_{\rm R-LS}}\left(\bx^{r_{j+1}}\right) \\
\leq f_{_{\rm R-LS}}\left(\bx^{r_{j}+1}\right) \leq g\left(\bx^{r_{j}+1}|\bx^{r_{j}}\right)\leq g\left(\bx|\bx^{r_{j}}\right)
\end{array}
\end{equation}
which implies $g'(\bz|\bz) \geq 0$ and $f'(\bz) \geq 0$ \cite{convergence}.
Hence, $\bz$ is the stationary point of $f_{_{\rm R-LS}}\left(\cdot\right)$ and therefore the proposed algorithm converges to the stationary point.  
 
\section{Numerical simulations and experiments}\label{sec:4}
In this section, we first discuss the Cramer-Rao Lower bound for the estimation problem in (\ref{eq:12}). Then we compare \textbf{SOLVIT} with an existing method by various computer simulations and experiment.
\subsection{Cramer-Rao Lower Bound}
CRLB is the lower bound on the variance of an unbiased estimator, i.e., $\textrm{cov}(\hat{\bx}) \geq \bJ^{-1}(\bx)$, where ${\bJ(\bx)}$ is the Fisher Information Matrix and is given by:
\begin{equation}\label{crlb}
\bJ(\bx) = \bH \textrm{cov}(\br,\br)^{-1}\bH ^{T}
\end{equation}
where $\br$ is a vector containing all the distinct noisy range - differences, $\bH = [\bh_{2,1},\bh_{m,1}\cdots\bh_{m,m-1}]^{T}$ and $\bh_{m,n}$ is given by:
\begin{equation}
\bh_{m,n} = \dfrac{ \bx- \by_{m}}{\|\bx - \by_{m}\|} - \dfrac{ \bx- \by_{n}}{\|\bx - \by_{n}\|}
\end{equation}
From (4), it can be seen that $r_{ij} = r_{i} - r_{j}={\|\bx - \by_{i}\|}_{2} +\varepsilon_{i}-{\|\bx - \by_{j}\|}_{2} -\varepsilon_{j}$ for $\{i, j = 1,2 \cdots, m\}$ and since $[\varepsilon_{i}]_{i=1}^{m}$ is a zero mean independent Gaussian random variable, the entries of covariance matrix of $\textrm{cov}(\br,\br)$ is given by:
\begin{equation}
\begin{array}{ll}
\textrm{cov}(r_{i,j},r_{k,l}) =\left\{ \begin{array}{ll}
\sigma^{2}_{r_{i}}+\sigma^{2}_{r_{j}}& \textrm{if}\quad i = k \: \textrm{and}\: j = l \\
\sigma^{2}_{r_{i}} & \textrm{if}\quad i = k \: \textrm{and}\: j \neq l \\
\sigma^{2}_{r_{j}} & \textrm{if}\quad i \neq k \:\textrm{and}\: j = l\\
-\sigma^{2}_{r_{i}} & \textrm{if}\quad i = l \:\textrm{and}\:  j \neq k\\
-\sigma^{2}_{r_{j}} & \textrm{if}\quad\: i \neq l \:\textrm{and} j = k\\
0& \textrm{if}\quad j\neq k\neq i \neq l
\end{array}\right..
\end{array}
\end{equation}
where $\sigma^{2}_{r_{i}}$ is the variance of the range $r_{i}$. The authors in \cite{bound_new}, computed the variance of the range estimation and is given by: 
\begin{equation}\label{variance}
\begin{array}{ll}
\sigma^{2}_{r_{i}}  = \dfrac{\sigma^{2}c^{2}{D_{i}}^{4}}{{\epsilon F_{s}}\left(c^{2}+ {D_{i}}^{2}{\bar{F^{2}}}\right)}
\end{array}
\end{equation}
where $c$ is the signal propagation speed, $D_{i} = {\|\bx - \by_{i}\|}_{2}$, $\epsilon$ is the transmitted signal energy, $\sigma^{2}$ is the variance of the additive noise at the output of the receiver, $F_{s}$ is the sampling frequency and $\bar{F}$ is the mean square bandwidth of the transmitted signal which is given by: 
\begin{equation}
\begin{array}{ll}
\bar{F^{2}}=\dfrac{\int_{-\infty}^{\infty} \left(2\pi F\right)^{2}|S(F)|^{2}dF}{\int_{-\infty}^{\infty} |S(F)|^{2}dF}
\end{array}
\end{equation}
where $S(F)$ is the Fourier transform of the transmitted signal. If the transmitted signal is taken to be a cosine signal with a frequency $f_{0}$ and time period $T_{0}$, then the transmitted signal energy $\epsilon$ is equal to  $\dfrac{T_{0}}{2}$ and the mean square bandwidth is equal to $4\pi^{2}{f_{0}}^{2}$. Substituting these values in (\ref{variance}) we get:
\begin{equation}\label{variance_sinusoidal}
\begin{array}{ll}
\sigma^{2}_{r_{i}}  = \dfrac{\sigma^{2}c^{2}{D_{i}}^{4}}{{2}\left(c^{2}+ {4\pi^{2}{f_{0}}^{2}}{D_{i}}^{2}\right)}
\end{array}
\end{equation}
where we have taken the sampling frequency $F_{s}$ to be $4f_{0}$. From (\ref{variance_sinusoidal}), it can be seen that the CRLB depends on the speed of the signal, noise variance and frequency of the transmitted signal.
\subsection{Numerical Simulations}
In this sub-section, we present numerical simulations to compare the proposed algorithm with the state-of-the-art positioning algorithms which estimate the source position from the range-difference measurements. In particular, we compare \textbf{SOLVIT} with the following methods - reference-free methods proposed in \cite{refreeold}, \cite{ono1} and \cite{amar} and a reference-based method proposed in \cite{beck}. For the reference-based method, the first sensor was chosen as the reference sensor. The iterative source localization algorithms were made to run either until the convergence criteria were satisfied or until the maximum number of iterations was reached. We do not show any simulations for algorithms on the R-LS problem as we have only rederived the SFP algorithm and numerical simulations regarding SFP can be found in \cite{SFP_SWLS}. The root mean square error is taken as the performance measure which is computed as:
\begin{equation}
\begin{array}{ll}
\textrm{RMSE} = \sqrt{\dfrac{1}{N_{exp}}{\displaystyle \sum_{i=1}^{N_{exp}}}{\|\bx -\hat{\bx}\|}_{2}^{2}}
\end{array}
\end{equation} 
where $\hat{\bx}$ is the estimated source position and $N_{exp}$ is the number of simulations which is taken as $500$. We compare the RMSE with that of the variance obtained using the Cramer-Rao Lower Bound (CRLB) in (\ref{crlb}). \\
\\
1. In this simulation we show that the proposed algorithm has monotonic convergence. The position of the sensors was randomly generated from a uniform distribution
from $[-10, 10]^{n}$, for $n=2$ and $m=4$. The source was assumed to be at $[10, 10]^{T}$. Fig. \ref{fig1} shows the monotonic behavior of \textbf{SOLVIT} for two different initialization - a random initialization from a uniform distribution from $[0, 1]^{n}$ and an initialization based on the scheme proposed in Section. \ref{initialize}.   
\begin{center}
\includegraphics[height=2.2in,width=3.5in]{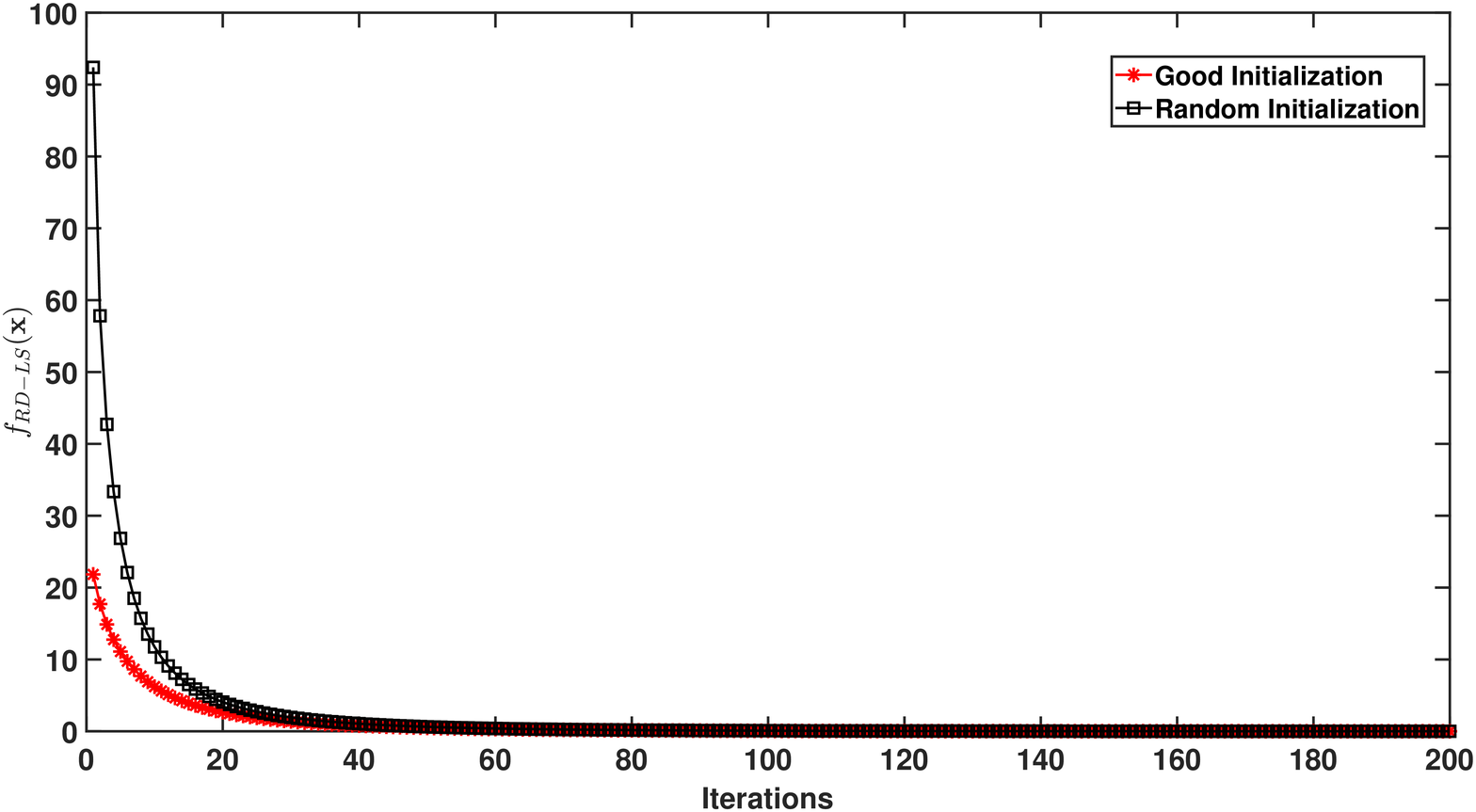}
\begin{figure}[!h]
\caption{Monotonic convergence of \textbf{SOLVIT} for $n=2$ and $m=4$. Red star line - Performance of \textbf{SOLVIT}  when initialized with the proposed scheme  in Section. \ref{initialize}; Black square line - performance of \textbf{SOLVIT} for random initialization.}
\label{fig1}
\end{figure}
\end{center}
From Fig. \ref{fig1} it can be seen that \textbf{SOLVIT} converges monotonically for both the initializations and when initialized with the scheme proposed in Section. \ref{initialize}, it converges faster.\\ 
\\ 
2. In this simulation, we varied the SNR from $-20$ to $0$ dB in steps of $2$ dB and evaluated the performance of the positioning algorithms for $n = 2$ and $m = 5$. The position of the sensors was randomly generated from a uniform distribution from $[-50, 50]^{n}$. The source position was also randomly generated from a uniform distribution from $[-10, 10]^{n}$. We assumed the speed of sound as $340\:m/s$ and the frequency of the signal was set at $1000$ Hz. Fig. \ref{fig2} shows the RMSE vs  SNR for the case $n = 2$. The initial value for the iterative algorithms i.e. \textbf{SOLVIT} and the method in \cite{ono1} were kept same. As can be seen from Fig. \ref{fig2}, the proposed algorithm has a lower positional error when compared to the available state-of-the-art methods. 
\begin{center}
\begin{figure}[!ht]
\includegraphics[height=2.2in,width=3.5in]{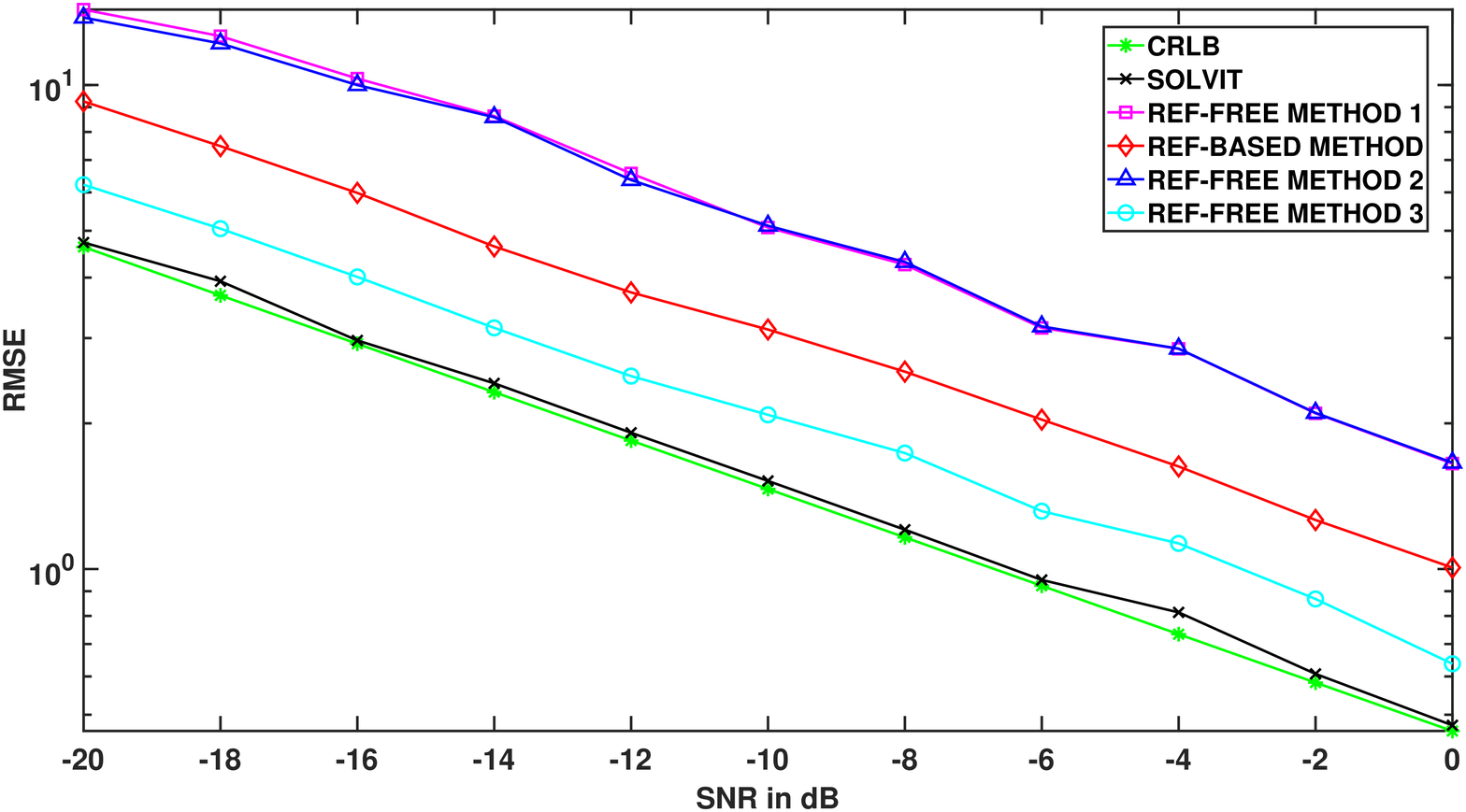}
\caption{RMSE vs SNR for array with random geometry for $n=2$ and $m=5$. 
Star green line - CRLB; ``x'' black line - \textbf{SOLVIT};  Diamond red line - reference-based method proposed in \cite{beck}; Circle cyan line - reference-free method proposed in \cite{ono1}; Triangular blue line - reference-free method proposed in \cite{refreeold}; Square magenta line - reference-free method proposed in \cite{amar}.}
\label{fig2}
\end{figure}
\end{center}
3. In this simulation, we consider arrays with deterministic geometry like circular, rhombus and linear array to evaluate the accuracy of positioning algorithms. The position of the $i^{th}$ sensor in the case of circular array was assumed to be at $\by_{i} = 10\left [\rm{cos} \left(\dfrac{2\pi  i}{m}\right), sin\left(\dfrac{2\pi  i}{m}\right)\right]^{T}$, where $m=6$. The position of the sensors for the rhombus array was assumed to be at  $\by_{1} = [0,10]^{T}$, $\by_{2} = [10,0]^{T}$, $\by_{3} = [0,-10]^{T}$ and $\by_{4} = [-10, 0]^{T}$. In the case of linear array, the position of the sensors were at $\by_{1} = [5,0]^{T}$, $\by_{2} = [5,10]^{T}$, $\by_{3} = [5,20]^{T}$ and $\by_{4} = [5,30]^{T}$. The position of the source for the circular and rhombus array was at $[1,5]^{T}$, while for linear array it was assumed to be at $[-5,5]^{T}$.  Fig. \ref{figg}(a), Fig. \ref{figg}(b) and Fig. \ref{figg}(c) shows the RMSE vs SNR for circular array, rhombus array and linear array, respectively.
\begin{figure}[!h]
\begin{subfigure}{0.49\textwidth}
\centering
\includegraphics[height=2.0in,width=3.5in]{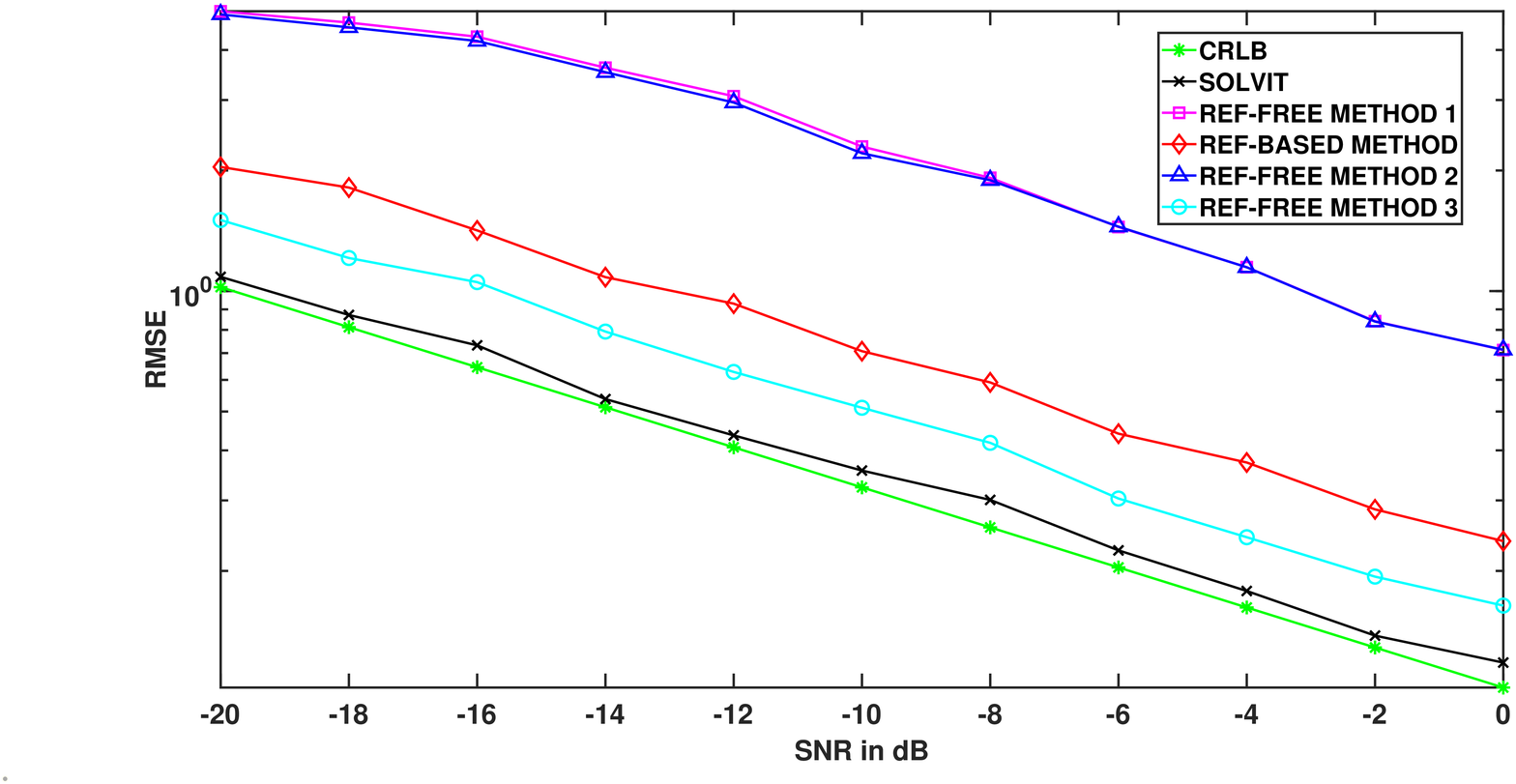}
\caption{RMSE vs SNR for circular array}
\end{subfigure}
\begin{subfigure}{0.49\textwidth}
\centering
\includegraphics[height=2.0in,width=3.5in]{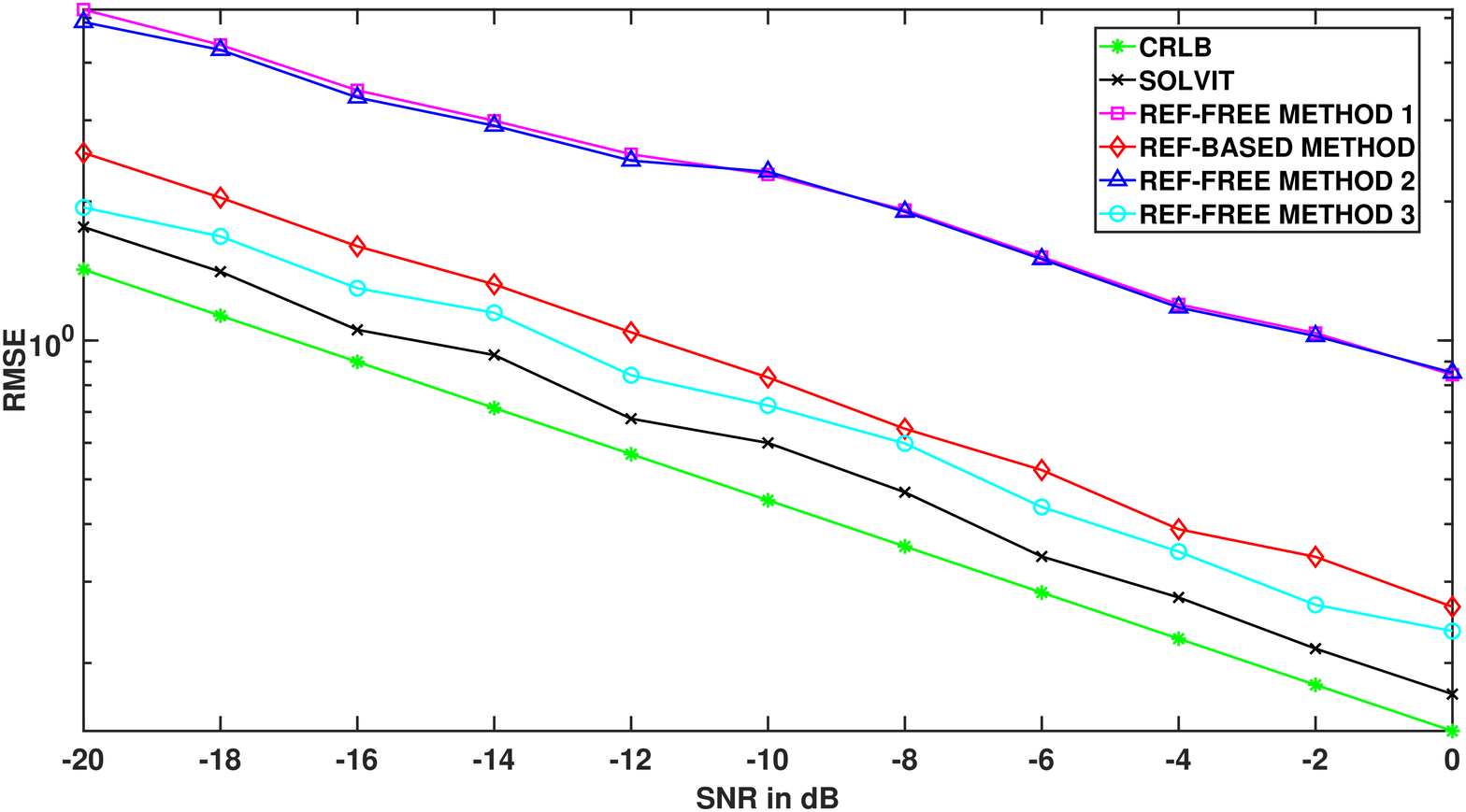}
\caption{RMSE vs SNR for rhombus array}
\end{subfigure}
\begin{subfigure}{0.49\textwidth}
\centering
\includegraphics[height=2.0in,width=3.5in]{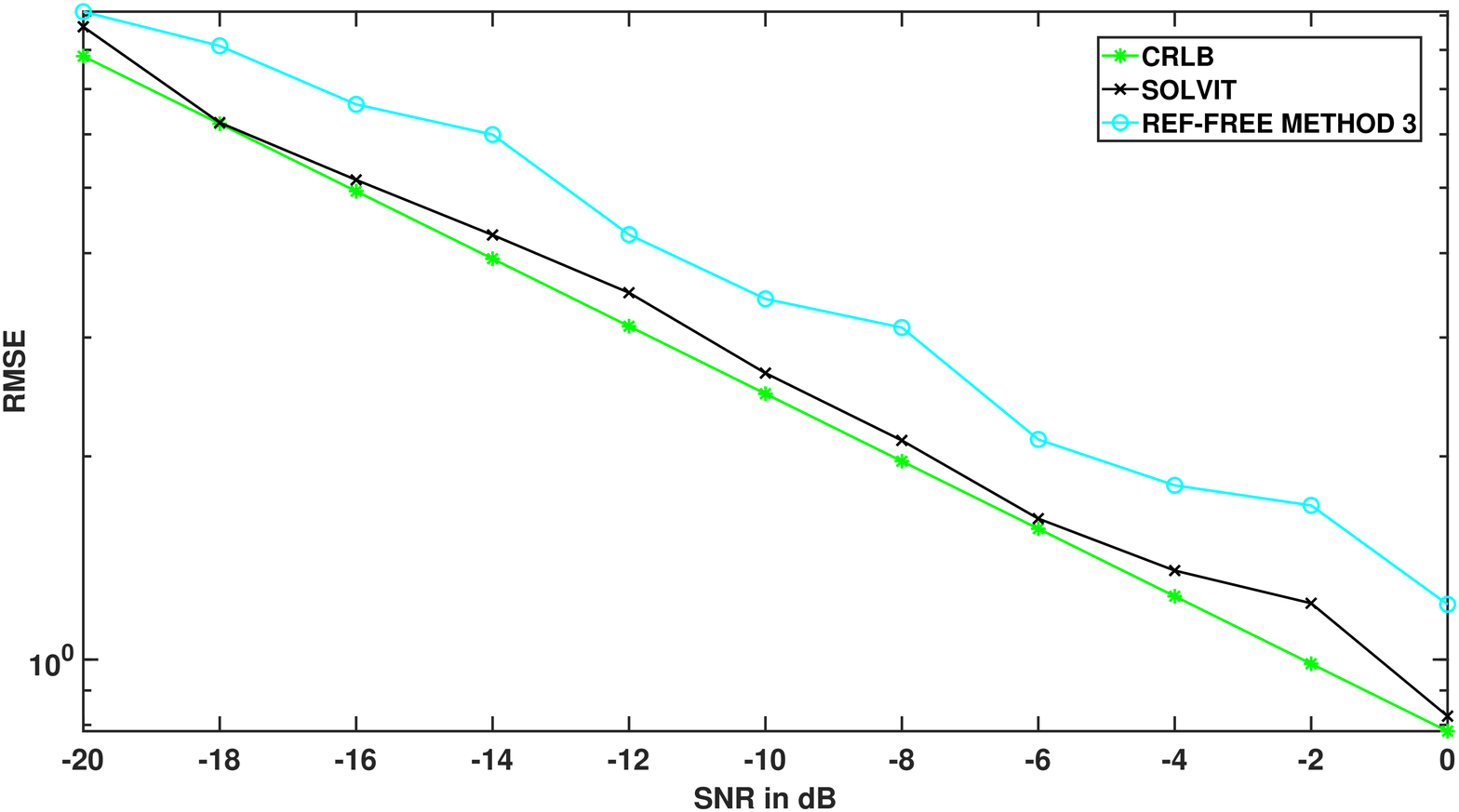}
\caption{RMSE vs SNR for linear array }
\end{subfigure}
\caption{Comparison of proposed algorithm with the reference-free method in \cite{amar} for various array geometries.
Star green line - CRLB; ``x'' black line - \textbf{SOLVIT}; Circle cyan line - reference-free method proposed in \cite{ono1}; Diamond red line - reference-based method proposed in \cite{beck}; Triangular blue line - reference-free method proposed in \cite{refreeold}; Square magenta line - reference-free method proposed in \cite{amar}}
\label{figg}
\end{figure}
From Fig. \ref{figg} it can be seen that the proposed algorithm performs better than the state-of-the-art reference-based and reference-free methods for all the three array geometries. Also, it was found that in the case of linear array, except for the method in \cite{ono1}, the other state-of-the-art methods becomes ill-conditioned and is unable to localize the source. However, as can be seen from Fig. \ref{figg}(c) \textbf{SOLVIT} is able to localize the source for linear array geometry with good positional accuracy. \\
\\
\noindent4. In this simulation we varied the frequency of the source signal from $100$ Hz to $10000$ Hz in steps of $1000$ Hz and evaluated the RMSE of the proposed algorithm with the method in \cite{amar}. A circular array geometry was considered with the position of the $i^{th}$ sensor as $\by_{i} = 10 \left [\rm{cos} \left(\dfrac{2\pi  i}{m}\right), sin\left(\dfrac{2\pi  i}{m}\right)\right]^{T}$, where $m=5$. The position of the source was assumed to be at $[1, 5]^{T}$, SNR was set at $0$ dB and the speed of the source was assumed as $340\:m/s$. The RMSE vs frequency of the source is shown in Fig. \ref{fig4}. and from the figure it can be seen that the proposed algorithm has lower RMSE for any frequency of the source signal, when compared to the state-of-the-art methods.
\begin{center}
\begin{figure}[!h]
\includegraphics[height=2.0in,width=3.5in]{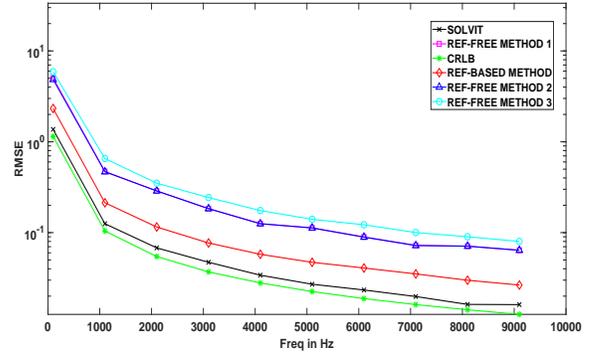}
\caption{RMSE vs frequency of the source for $n=2$ and $m=5$.
Star green line - CRLB; ``x'' black line - \textbf{SOLVIT};  Circle cyan line - reference-free method proposed in \cite{ono1}; Diamond red line - reference-based method proposed in \cite{beck}; Triangular blue line - reference-free method proposed in \cite{refreeold}; Square magenta line - reference-free method proposed in \cite{amar}.}
\label{fig4}
\end{figure}
\end{center}
\subsection{Real-life experiment}
\begin{figure}[H]
\centering
\begin{tabular}{c}
\includegraphics[height=2.0in,width=3.3in]{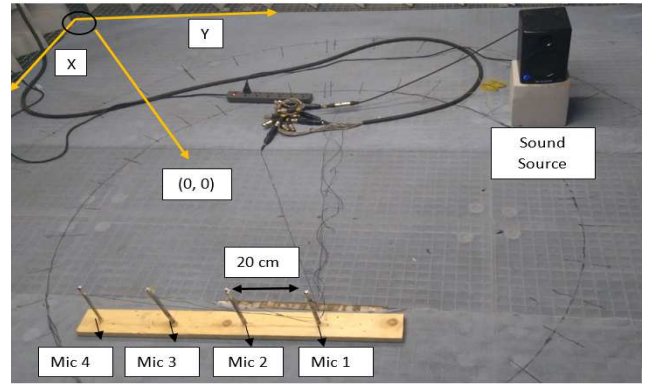}
\end{tabular}
\caption{Test setup inside an anechoic chamber}
\label{anechoic}
\end{figure}

We next evaluate the positioning accuracy of  \textbf{SOLVIT} in a real-life environment. To do so, experiments were conducted in an anechoic chamber. The length and height of the chamber are $3.340$ m and $3.040$ m, respectively. The experimental set up is as shown in Fig. \ref{anechoic}. A corner of the room was taken as the origin. A linear array with ${m =4}$ sensors was considered. The spacing between each sensor is ${20}$ cm. The sound source to be localized emits a sinusoidal wave of $250$ Hz ($100$ KHz as sampling frequency); which was sent via LabVIEW from a PC with 3.60 GHz Processor with 16 GB RAM  through a 24 channel, 2.4 mA PCI. The signal was received through LabVIEW from PCI which was further processed offline in a separate PC with 2.0 GHz Processor with 64 GB RAM in MATLAB. An IIR Bandpass filter first filtered the signal received by each microphone with a lower and higher cut-off frequency set as $150$ Hz and $350$ Hz respectively. The time delay between the ${i}$th and the ${j}$th microphone was then estimated using the standard cross-correlation technique. The estimated time delay is multiplied by the speed of sound in air to obtain the range-differences $r_{ij}$  $\{i, j  = 1,2\cdots 4\}$. The distinct range-differences was then given as input to \textbf{SOLVIT}. 
The four microphones were kept at $\by_{1} = [2.1, 1.7]^{T}$, $\by_{2} = [2.1, 1.5]^{T}$, $\by_{3} = [2.1, 1.3]^{T}$ and $\by_{4} = [2.1, 1.1]^{T}$. The mean square error of \textbf{SOLVIT} and the method in \cite{ono1} was found to be $0.04$ m and $9.55$ m, respectively. The mean square error along X component using \textbf{SOLVIT} and the method in \cite{ono1} was found to be $0.03$ m and $7.44$ m, respectively. The mean square error along Y component using \textbf{SOLVIT} and the method in \cite{ono1} was found to be $0.01$ m and $2.11$ m, respectively.  \\


\begin{table*}[!htbp]
\centering
\caption{True position and the Estimated position by \textbf{SOLVIT} and the method in \cite{ono1}.}
\label{table:2}
\begin{tabular}{|p{1cm}||p{1cm}|p{1cm}|p{2cm}|p{2cm}|p{2cm}|p{2cm}|}
\hline
 Source &X &Y &X &X&Y&Y\\
           &(True)&(True)&(Estimate by \textbf{SOLVIT})&(Estimate by \cite{ono1})&(Estimate by \textbf{SOLVIT})&(Estimate by \cite{ono1})\\
 \hline
1   &   0.4 m   &  2 m & 0.16 m &0.1 m&2.1 m& 2.1 m \\
2   &   1 m   & 0.5 m  & 1.1 m& 1.16 m&0.5 m&0.55 m \\
3   &  1.35 m   &  0.7 m & 1.5 m  &1.6 m& 0.8 m&0.86 m \\
4   &    0.4 m &  1 m & 0.15 m &-4.6 m&0.9 m&-0.25 m\\
5  &   0.5 m   & 1.9 m & 0.27  m  & 0.17 m&2.1 m &2.1 m\\
6   &   0.4 m    &  0.4 m  &0.32 m&-4.76 m  & 0.21 m&-3.22 m \\
7   &    1 m & 2.5 m  & 0.95 m &0.53 m& 2.4 m&2.78 m\\
\hline
\end{tabular}
\end{table*}

\section{Conclusion}\label{sec:5}
In this paper, we proposed an MM based algorithm - \textbf{SOLVIT}; which solves the RD-LS problem and thereby localizes the source without making any assumption on the reference sensor and without squaring the range-difference measurements. We also prove that \textbf{SOLVIT} is monotonic and converges to the stationary point of the RD-LS problem. Various numerical simulations were done to evaluate the positioning accuracy of \textbf{SOLVIT}. It was evaluated in various sensor array geometries and it was found that while some of the existing reference-free and reference-based methods become ill-conditioned for certain source and sensor positions; \textbf{SOLVIT} at least returns a local optimal solution to the problem. \textbf{SOLVIT} was also evaluated for varying source frequencies and it was found that it performs better than the existing state-of-the-art algorithms in terms of positioning accuracy. Experiments were conducted in the anechoic chamber to evaluate the performance of the algorithm in a real-world scenario. The mean square error was found to be $0.04$ m. Hence, from the simulations and experiments we conclude that \textbf{SOLVIT} has better positioning accuracy when compared to the available methods. We also solve the R-LS problem and rederive the SFP algorithm using MM and show a less intricate proof of convergence.\\
\bibliographystyle{IEEEtran} 
\bibliography{refs}
\end{document}